\documentclass[aps,prd,preprint,groupedaddress]{revtex4}

\usepackage{amsmath,amsthm,amscd,latexsym,amssymb}
\usepackage{graphicx}

\newcommand{\C}{\mathcal C}

\def\d{\partial}
\def\p{\varrho}
\def\w{\omega}
\def\Vac{\Omega}

\def\eps{\epsilon}

\def\a{\alpha}
\def\be{\beta}

\def\w{\omega}
\def\Vac{\Omega}

\def\s{\sigma}

\def\vp{\varphi}
\def\la{\lambda}

\def\h{{\mathsf h}}
\def\H{\mathcal H}
\def\Rup{\overrightarrow{\mathsf R}}
\def\Rin{\overleftarrow{\mathsf R}}
\def\Bin{\overleftarrow{\mathsf B}}
\def\Rpl{{\mathsf R}_{lp}}
\def\Rwl{{\mathsf R}_{l\w}}
\def\R{r_s}

\def\dens#1{\langle\varrho(#1)\rangle}

\def\V{\mathcal V}

\newsymbol\rest 1316

\def\d{\partial}
\def\D{\nabla}

\def\*{\cdot }

\def\ra{\rightarrow}
\def\rest{\upharpoonright}

\def\C{\mathcal{C}}

\def\Fou{\mathcal{F}}

\def\ess_sp{\s_{ess}}
\def\F1{\Fou^{(1)}}

\def\C{\mathcal C}

\def\dom{\C_0^\infty}

\newtheorem{theorem}{Theorem}
\newtheorem{lemma}{Lemma}

\usepackage{hyperref}

\begin{document}


\title{Bounds on the energy densities of ground states on static spacetimes of compact objects}


\author{Piotr Marecki}
\email[]{piotrm@wsi.edu.pl}
\affiliation{Wyzsza Szkola Informatyki i Zarzadzania,
Bielsko-Biala, Poland}


\date{\today}

\begin{abstract}
In this paper we investigate quantum fields propagating on given,
static, spherically symmetric spacetimes, which are isometric to a
part of the Schwarzschild spacetime. Without specifying the
internal geometry we show, that there exist bounds on the energy
densities of ground states of a quantum scalar field on such
spacetimes. The bounds (from above and below) come from the
so-called Quantum Energy Inequalities, and are centered around the
energy density of the Boulware state (the ground state for
Schwarzschild spacetime). The specific value of the bound from
below depends critically on the distance $\ell$ from the horizon,
where the spacetimes of compact objects cease to be isometric to
the Schwarzschild spacetime. In the limit of small $\ell$ we
prove, that the energy densities of ground states cannot be below
the Boulware level.
\end{abstract}

\pacs{}

\maketitle


\section{Introduction}

Quantum fields living on a given spacetime can violate the
classical energy conditions, in particular, some states can
exhibit negative energy densities. However, energy density
operators (which are classically non-linear in fields) need to be
constructed in a local manner \cite{BFV}, which makes them
invariant to the changes of geometry outside of the region of
their support. This means, that the two-point functions of quantum
states cannot be employed in the definition of pointwise products
of fields, as is the case for normal-ordering. Thus, even the
``vacuum states'' exhibit non-zero expectation values of the
energy-density operator. On a generic globally hyperbolic
spacetime there exists at present no condition to distinguish a
preferred vacuum state, apart from the necessity for it to be a
Hadamard state (only for these states we know a systematic
construction of operators non-linear in fields). For static
spacetimes, however, ground states are distinguished and
acceptable, at least if infrared problems do not prohibit their
existence. The energy densities of ground states have been
calculated, particularly for the spacetimes possessing horizons
(eg. the Schwarzschild spacetime) or time-like boundaries
\cite{Cand,Cand2}. Generally, these densities tend to infinity (in
absolute) if the proper distance form the boundary, denoted here
by $L$, approaches zero. For spacetimes without
boundaries/horizons the densities are everywhere finite (because
the ground states are Hadamard states \cite{SV}), although it is
difficult to obtain concrete values of these important quantities.
In this paper we propose a method to put upper and lower bounds on
the energy densities of ground states for spacetimes of compact
objects. The geometry of these spacetimes is assumed to be
horizon-free and isometric to the Schwarzschild geometry for
points at the proper distance $\ell$ and larger from the
Schwarzschild horizon. In other words: we consider a class of
spacetimes parameterized by $\ell$, but leave the internal
geometry of these spacetimes unspecified. We will employ Quantum
Energy Inequalities \cite{Ford,FordRoman} to derive the bounds on
the energy densities of ground states. Let us introduce the QEI,
as is usually done, on the example of massless fileds in the
Minkowski spacetime. In this case the energy density of the ground
state (the Poincare-invariant vacuum) is zero, and for an
arbitrary state $\psi$ the QEI gives
\begin{equation}\label{int}
    \langle \p \rangle_\psi\geqslant -\frac{3}{32 \pi^2 T^4}
\end{equation}
for an energy density operator smeared in time with the Lorentzian
test function
\begin{equation}
w^2(t)=\frac{T}{\pi(T^2+t^2)},
\end{equation}
(the reason for the notation will later become clear). If we
interpret $T$ as the characteristic time scale of  measurement, we
van infere from \eqref{int}, that periods of negative energy
density fulfill a uncertainty-like inequality restricting their
duration and magnitude. Quantum energy inequalities by now have
become a reliable tool in quantum field theory: they have been
proved for arbitrary test functions \cite{Few}, also on curved
spacetimes with respect to arbitrary reference Hadamard states
(difference quantum inequalities of \cite{Few_gen}). Also, these
results are not specific to scalar fields: there are corresponding
QEIs for the Dirac (eg. \cite{FV,FM}) and electromagnetic (eg.
\cite{FordRoman,FP1}) fields.

In this paper we follow a line of thought proposed  by Fewster and
Pfenning \cite{FP}, namely, for static spacetimes isometric in a
certain causally complete region: on the one hand we can identify
the local observables supported in this region (\cite{BFV}), and
on the other we can restrict the ground states of both spacetimes
to this region. By developing QEIs with either of these two states
as the reference state we obtain two inequalities which are
necessarily fulfilled by the difference of the energy densities of
these states. Thus, if the energy-density of one of these states
is known, the other will be bounded from above and below by the
QEIs. In section III.1 we make an important geometrical
observation, which tells us that although for static observers
located in the exterior Schwarzschild region there exists an upper
bound on the maximal sampling time $T$, its value can still become
very large if the compact object is described by Schwarzschild
geometry up to a very small distance from the Schwarzschild
horizon. Therefore, even for small distances $L$ from the horizon
there can still be long sampling times $T$, provided that $\ell$
is small enough.

The paper is organized as follows: the second section contains all
 preliminaries necessary for a treatment of massless quantum
fields in static spacetimes, as well as the appropriate version of
quantum energy inequalities. In the third section we describe
briefly the geometry of spacetimes under consideration, with
special emphasis on the maximal size of causally complete regions
(double-cones) which can be fit into the region isometric to the
Schwarzschild spacetime. The fourth section is the main section of
this paper. It contains a description of the method which we have
outlined above, as well as its application to the problem of
finding bounds on the energy density of ground states of compact
objects. We develop bounds from below and above separately. In the
former case, we prove a theorem (theorem 1) which restricts the
energy density of the ground state from below (by anchoring it
against the energy-density of the Boulware state). In the limit of
very small $\ell$ this bound effectively tells us, that the energy
density of ground states cannot be much lower than that of the
Boulware state. Our results on the bound from above are weaker: we
prove, that a bound form above on the energy density of ground
states exists, although we cannot guarantee at present, that this
density cannot be much higher than that of Boulware state in the
limit of very small $\ell$ (i.e. if the compact objects ``tend
to'' a black hole). It should be stressed, that the Boulware state
appears here only as a technical tool (convenient, because the
two-point function and the energy density for this state are
known), employed in the derivation of bounds on the energy density
of ground states of spacetimes without horizons.

Obviously, all these results are valid only if the ground states
considered here exist. This is the case at least for spacetimes
fulfilling the condition \eqref{condition} of appendix B.

\section{Quantum fields in Curved Spacetimes and Quantum Weak Energy Inequalities}

\subsection{Preliminaries}\label{pre}

In this paper we use the Landau-Lifshitz time-like convention for
the metric, the Riemann and the energy-momentum
tensors\footnote{This means we use the convention $-++$ in the
classification of Misner, Thorne and Wheeler\cite{MTW}.}. We will
consider
 static, spherically symmetric spacetimes, with the metric:
\begin{equation}\label{metric}
  ds^2=f(r) dt^2-\frac{dr^2}{h(r)} -r^2 (d\theta^2 +\sin^2\theta d\phi).
\end{equation}
We will consider (globally hyperbolic) spacetimes of compact
objects, for which $r\in[0,\infty)$ and $f(r), h(r)$ are strictly
positive. Additionally, we will also consider the Schwarzschild
spacetime, where $r\in (\R,\infty)$, $f(r)=(1-\R/r)=h(r)$. ($\R$
stands here for the Schwarzschild radius $\R=2M$.) We will
frequently use the abbreviation $g(r)=\sqrt{fh}$. The surfaces of
constant $t$ are Cauchy surfaces for spacetimes which we consider;
these surfaces will be denoted by $\Sigma_t$.

We will now review the standard structure \cite{Wald_CST}, adapted
to the spacetimes we consider, associated with the evolution of
classical scalar field $\phi$. Let $S$ denote the space of real
initial data, compactly supported on $\Sigma_t$:
\begin{equation}
  S=\dom(\Sigma_t)\oplus \dom(\Sigma_t).
\end{equation}
The minimally coupled, massless scalar field, fulfills the
equation
\begin{equation}
\Box \phi(x)=0,
\end{equation}
which can be rewritten   in the form
\begin{equation}
  \frac{\d^2 \phi}{\d t^2}=-A \phi.
\end{equation}
In our case
\begin{equation}
 -A u=-f\D^i\D_i u=\frac{g}{r^2}\d_r\left[g r^2 \d_r u\right]+\frac{f\hat
  L^2}{r^2} u,\quad {\text{for\ }u\in \mathcal D(A),}
\end{equation}
where $\D^i$ denotes the 4D covariant derivative taken for spatial
indices $i=(r,\theta,\phi)$ and $\hat L^2$ stands, as usual, for
the squared angular momentum operator. Introducing the
Regge-Wheeler-like coordinate ``$x$'', $\d_x=g\d_r$, we make use
of the identity
\begin{equation}
\frac{g}{r^2}\d_r\left[g r^2 \d_ru\right]=\frac{1}{r}\left[\d^2_x(r u) -
g (\d_r g)u \right].
\end{equation}
Consequently, the operator $A$ acquires the form
\begin{equation}
A(ru)=\frac{1}{r}\left[-\d^2_x  +\frac{g\d_r g}{r}  - \frac{f\hat
L^2}{r^2}\right] (r u),
\end{equation}
$A$ is initially defined on the dense domain of smooth functions
compactly supported on the Cauchy surface (these space of
functions will be denoted by $\dom(\Sigma_t)$), on which we
introduce the scalar product
\begin{equation} \label{sc}
(u_1,u_2)=\int_{\Sigma_t} \overline{u_1(\vec x)} u_2(\vec x)\
\frac{r^2\, drd\Vac}{g(r)},
\end{equation}
which is the natural (Lebesgue) measure\footnote{This measure is
not the proper volume on $\Sigma_t$. It is the measure derived
from the Gauss's theorem relating 4D integrals of 4-divergencies
of vector fields to 3D integrals over the Cauchy surfaces of
constant $t$ (for spatially compactly supported vector fields).}
on $\Sigma_t$ . (Here $d\Vac$ denotes the volume of the 2-sphere.)
The complex Hilbert space of square integrable functions w.r.t.
the above scalar product will be denoted by $\H$. This space is
often called the one-particle Hilbert space.

We will investigate only these spacetimes, where $A$ is a positive
operator (see appendix \ref{KMS}); in this case there exists a
square root of $A$, the `one-particle Hamiltonian', denoted by
$\h$. The dynamics of classical fields, and the construction of
ground-state representation of the quantum field algebra for such
spacetimes is standard \cite{Wald_CST,AH}; we will recall the most
important features of it for the convenience of the reader.

Let $\Phi=(v,v_t)\in S$ denote some arbitrary initial data. Then the
transformation $\V(t): \Phi\ra \Phi(t)$
\begin{align}
  v(t)= & \cos (\h t) v + \sin(\h t) \h^{-1} v_t  \\
  v_t(t)= &-\sin(\h t)\h v +\cos(\h t) v_t,
\end{align}
describes the temporal evolution of these data. The symplectic
form
\begin{equation}
  \sigma (\Phi_1,\Phi_2)=\int_{\Sigma_t}\left[\Phi_1  \D_a \Phi_2-(\D_a \Phi_1) \Phi_2\right]n^a d\eta=  \int_{\Sigma_t} (v_1 v_{t2} - v_{t1} v_2)
  \frac{r^2\, dr d\Vac}{g(r)},
\end{equation}
(where the unit vector $n^a$ is the normalized version of
$(\d_t)^a$, and $d\eta$ is the volume element on $\Sigma_t$) is
preserved by $\V(t)$
\begin{equation}
  \s \bigl(\V(t) \Phi_1,\V(t) \Phi_2\bigr)=\s (\Phi_1,\Phi_2)
\end{equation}

In order to construct the ground state, one introduces an operator
$k:S\ra \H$, which extracts the positive frequency part of the
Cauchy data; for $[v,v_t]\in S$ :
\begin{equation}
    k [v,v_t]=\frac{1}{\sqrt{2}}\left[i \sqrt{h}\, v + \frac{1}{\sqrt{h}}\, v_t\right]
\end{equation}
The classical structure, that is the one-particle structure, $(\H,
k,e^{i\h t})$, and the symplectic space $(S,\s, \V(t))$ can be
quantized in the usual manner:  one constructs the standard Fock
space, and introduces the creation and annihilation operators,
$a,a^*$. In this way the ground-state representation of the
algebra of free fields is obtained: for a real test function
$\chi$ the smeared field operator is obtained from
\begin{equation}
\vp(\chi)=a(k[0,\chi])+a^*(k[0,\chi])
\end{equation}
and its time derivative from
\begin{equation}
\dot \vp(\chi)=a(k[\chi,0])+a^*(k[\chi,0]).
\end{equation}

Some contact can be made with the more common expressions, namely,
according to the spectral theorem, there will be a spectral
measure associated with the (self-adjoint) operator $A$, $d\mu_I$,
which allows for a functional calculus w.r.t. $A$ ($I$ will index
the generalized eigenfunctions of $A$ (the modes), denoted by
$F_I(\vec x)$, corresponding to the eigenvalues $\w_I$ (the
frequencies)). Than we may express the operator-valued
distribution $\vp(t,\vec x)$ as
\begin{equation}
    \vp(t, \vec x)=\int d\mu_I\, \frac{1}{\sqrt{2p}} \left[e^{i\w_I t} F_I(\vec x)\, a_I +
    e^{-i\w_I t} \overline F_I(\vec x)\, a^*_I\right]
\end{equation}

\subsection{QEI for static spacetimes}

Here we shall review some of the results of \cite{Few}, where
Quantum Energy Inequalities have been proved for scalar quantum
fields propagating on static spacetimes.

Consider a static observer, located at the spatial position $\vec
x$, whose (normalized) tangent vector is $u^a$. Denote by
$\rho(w,\vec x)$ the (smeared) energy density operator which, we
assume, has been constructed in a local and generally
co\-va\-riant manner\cite{BFV}, with the help of the
point-splitting procedure employing a local Hadamard parametrix.
The smearing is such, that it corresponds to a measurement
performed by the static observer, sampled in time by a smooth,
non-negative function $w^2(t)$:
\begin{equation}\label{ro}
  \rho(w,\vec x)\doteq \int dt\ T_{ab}(\vec x,t)\ u^a u^b\, w^2(t).
\end{equation}
The sampling function is normalized as a probability density:
\begin{equation}
\int dt\ w^2(t) =1.
\end{equation}
 Fewster and Teo  have shown  \cite{Few}, that the energy-density operator, $\p(w,\vec
x)$, necessarily fulfills the inequality:
\begin{equation}\label{QEI}
   \langle\p(w,\vec x)\rangle_\psi - \langle \p(w,\vec x)\rangle_G
\geqslant Q_G[w,\vec x],
\end{equation}
 where $G$ is the ground state, $\psi$ any other state in the folium of $G$ (i.e.
 $\psi$ need not be pure, and can be prescribed by an arbitrary `density matrix'), and
\begin{equation}
 Q_G[w,\vec x]\doteq-\frac{1}{\pi}\int_0^\infty
  d\w
  \int \frac{d\mu_I}{2\w_I}\left(\frac{\w_I^2}{f(t)}-\frac{1}{4}\D_i \D^i
  \right)|F_I(\vec x)|^2\*g(\w+\w_I).
\end{equation}
(We hope $g(\w)$ will not be confused with $g(r)$ which
parameterizes the metric.) The inequality holds for all smooth
compactly supported sampling functions\footnote{In \cite{Few} a
more general family of functions was allowed (smooth functions,
decaying as $O(t^{-2})$ for large $t$). Here we will confine
ourselves to functions of compact support.} $w(t)$, which are
related to $g(\w)$, by $g(\w)=|\widehat{w}(\w)|^2$ (the `hat'
stands for the Fourier transform), that is
\begin{equation}\label{def_g}
g(\w)=\left| \int_{-\infty}^\infty dt\, w(t) e^{-i\w t} \right|^2.
\end{equation}
Note, that this function is rapidly decaying for large $\w$,
because $w(t)$ is smooth and compactly supported.

 We emphasize, that the right-hand-side of the
inequality \eqref{QEI} (denoted in the sequel by $Q_G[w,\vec x]$,
because it is a functional of $w$, constructed relatively to the
ground state $G$) is negative\footnote{The sign of the integrand
is positive for all $\vec x$ and $\w$ and $\w_I$, as can readily
be verified with the help of the wave equation.} and finite
whenever the ground state $G$ exists. This expresses the basic
message of QEIs, that there is a meaningful restriction on the
sub-ground energy density, which depends on $G$ and $w(t)$ alone,
for all states of the quantum field.

We will now adapt the QEI to the spherically symmetric case. The
angular dependence of the generalized eigenfunctions of $A$ can be
separated in the usual way,
\begin{equation}
  F_I(\vec x)=\frac{1}{\sqrt{2\pi }}\, \Rwl(r)\
  Y_{lm}(\theta,\phi).
\end{equation}
which with the help of
\begin{equation}
  \sum_{m=-l}^l |Y_{lm}|^2=(2l+1)/4\pi,
\end{equation}
leads to a further simplification:
\begin{multline}\label{QEIss} Q_G[w,\vec x]=
-\frac{1}{4\pi^2 f(r)}\int_0^\infty
  d\w\int_0^\infty \frac{dp}{p}
  \left(p^2-\frac{A}{4}
  \right) \left[\sum_{l=0}^\infty (2l+1) |\Rpl(r)|^2\right]\*g(\w+p),
\end{multline}
We stress, that the differential operator in the round brackets is
such, that the integrand is point-wise positive, for all $r,\w$
and $p$. The square bracket contains what is usually called ``a
mode sum''. This sum is typically difficult to evaluate, even
approximately, in concrete models. However, as we shall see in the
case of ``long measurements'' (specified below) only the
low-frequency behavior of this sum will be relevant. In some
cases, notably in the case of the Schwarzschild spacetime in the
region close to the horizon, this behavior is known explicitly. In
section \ref{bounds} we will use \eqref{QEIss} to derive upper and
lower bounds on  $\langle \varrho(r)\rangle_G$, for the ground
state $G$ on the spacetime of a compact object. For such
spacetimes, it would require a careful analysis to calculate the
mode sums for arbitrary $\w$ \footnote{These mode sums are
non-local, i.e. their value at a certain point depends on the
metric on the \emph{whole} Cauchy surface. In our concrete case,
the propagation of waves through the matter of the object needs to
be considered, even in the calculation of the sums in the exterior
region.}. Nonetheless, general results settle the $\w$-dependence
of these sums at small frequencies (see appendix \ref{KMS}), and
these will employed to derive bounds on the energy density
$\dens{r}_G$.

\section{Geometric observation}\label{observ}
\begin{figure}[h]\centering
\includegraphics[scale=0.8]{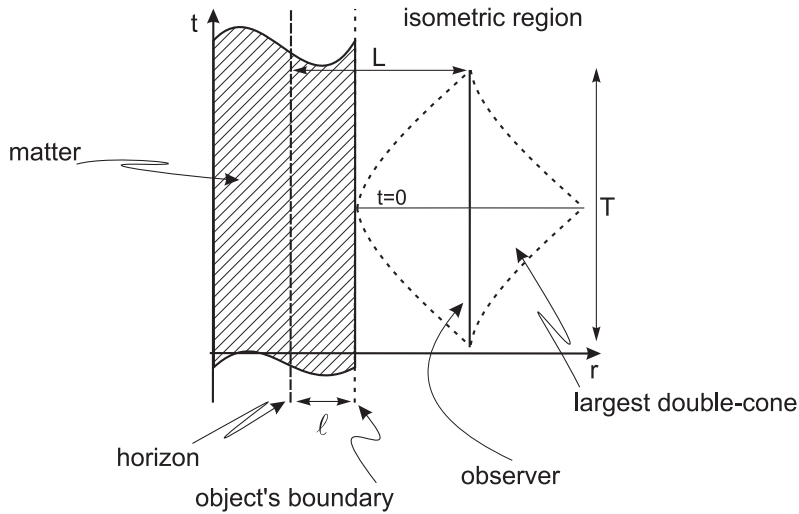}
\caption{Geometrical setting.\label{shell} The exterior part of
the spacetime of a spherical object is isometric to the
Schwarzschild spacetime (the object is compact).}
\end{figure}

We will now specify the geometry under consideration. Let us take
a Cauchy surface of constant $t$ in the Schwarzschild spacetime
and draw a sphere at the geodesic distance $\ell$ from the
horizon. The interior of this sphere will be filled with matter of
some sort (we do not wish to make any choice in  this paper, but
note that spacetimes of this sort have been constructed
\cite{MM,VW}). The situation with very small $\ell$ will be of
particular interest for us, as it corresponds to a static
spacetime which is ``on the verge'' of collapse. We will also
consider a static observer, located at the proper distance $L$
(measured along the radial geodesic) from the horizon (see fig.
\ref{shell}).
We introduce two dimensionless quantities
\begin{align}
    \eps_1&=a/\R -1,\\
    \eps_2&=b/\R -1,
\end{align}
where $a$ is the radial coordinate of the surface of the object,
and $b$ is the radial coordinate of the observer. It is a simple
task to find the largest double-cone contained in the isometric
region of both spacetimes. This in turn will give us the longest
duration of common measurements allowed by locality arguments (see
appendix \ref{local}). We set up a radial, outgoing null geodesic,
crossing $t=0$ at $r=a$, and search for the intersection with the
word-line of the observer. The longest available time of
measurement, found in this way is
\begin{equation}\label{long}
T=2 \R\left[\eps_2-\eps_1 +\ln(\eps_2/\eps_1)\right].
\end{equation}
Moreover, for $\eps_1<\eps_2\ll 1$ a simple connection between
$\eps_1,\eps_2$ and $\ell,L$ exists:
\begin{align} \label{a_rela}
    \eps_1&=(\ell/2\R)^2,\\
    \eps_2&=(L/2\R)^2.
\end{align}
For very small $\ell$ and $L$, the equation \eqref{long} still
allows for a fairly long measurement times $T$ if, for instance,
$\ell$ and $L$ differ by one order of magnitude. The largest
double-cone, spanned by an observer at $L$ has a duration
(measured in the global time) of
\begin{equation}\label{T}
    T\approx {4 \R} \ln(L/\ell).
\end{equation}
We emphasize that there is an essential difference between $T$
derived here and (for instance) $T$'s derived for static observers
in a flat spacetime. There, if we imagine a static boundary
located at $x=\ell$ and an observer at $x=L$, then the longest
time of measurement for which an analogous double-cone still does
not ``touch'' the boundary is $T=L-\ell$. Thus making $\ell$ small
does not prolong $T$ very much - in contrast to the result in our
setting, eq. \eqref{T}.

\section{Bounds on the energy density of ground states for static spacetimes of compact objects}
\label{bounds}
\subsection{Scaling and the bounds for long
measurements}\label{scaling} For the purposes of the argument
presented in this paper, it is necessary that the smearing
function, $w(t)$ be of compact support. In order to investigate
the limit of long measurements, we shall chose a  certain function
$w(t)$, and consider its normalized, scaled version:
\begin{equation*}
  w_\la(t)=\sqrt \la\ w(\la t).
\end{equation*}
The scaling factor is chosen in such a way, that the rescaled
sampling functions remain properly normalized, $\int dt\,
w^2_\la(t)=1$. Evidently, the limit of small $\la$ will correspond
to long measurements. The rescaled $g_\la(\w)$ will be important
in what follows,
\begin{equation*}
  g_\la(\w)=\frac{1}{\la}g(\w/\la).
\end{equation*}

\subsection{Bounds on the difference of energy densities}
The main aim of this paper is to compare the (known) energy
density of the Boulware state $\dens{r}_B$ (defined on the
Schwarzschild spacetime) with the unknown energy densities of
ground states $\dens{r}_G$ for spacetimes of compact objects. The
comparison will be made at the points lying in the region, where
both spacetimes are described by the Schwarzschild metric.

However, two different geometries lead to inequivalent
\emph{global} algebras of observables, which makes it conceptually
difficult to compare physical predictions, such as expectation
values of certain global observables (eg. the total energy, cf.
\cite{AH}). For purposes of this paper, we will need to compare
the expectation values of a quantum observable (the energy
density) in different states (which are defined on different
spacetimes); it is necessary to verify, that we take \emph{the
same} observable in both cases (it does not suffice, for instance,
if the geometry in the vicinity of the world-line of the measuring
apparatus is identical, see appendix \ref{local}). A sort of
comparison we need is possible only  for causally complete regions
with compact closure, such as double-cones of the form
$\mathcal{O}=J^+(x)\cap J^-(y)$, if such  regions can be
isometrically embedded into both spacetimes (fig. \ref{shell}).
Then, the algebras of local non-linear observables $\mathcal
W(\mathcal O)$ associated with this region (that is, algebras of
free fields, Wick polynomials, time-ordered products and
energy-momentum-tensor operators smeared with test functions
supported in the region $\mathcal O$) will be
isomorphic\cite{BFV}. This allows us to select a single local
observable, the energy density smeared with a sampling function
supported in $\mathcal{O}$, $\rho(w,\vec x)$, and compare its
expectation values in the states $B$ and $G$. Both of these states
are ground states for the respective static spacetimes.

Suppose we take the QEI \eqref{QEI} with respect to the Boulware
state $B$. It follows, that for any Hadamard state $\psi$ (on the
algebra of observables on the Schwarzschild spacetime) there
holds:
\begin{equation}
   \langle \p(w,\vec x)\rangle_\psi - \langle \p(w,\vec x)\rangle_B \geqslant Q_B[w,\vec
   x].
\end{equation}
If we now take the sampling function $w(t)$ with a compact
support, such that $\p(w,\vec x)$ belongs to the algebra of
observables, $\mathcal W(\mathcal O)$, of a double cone $\mathcal
O$ located in the region isometric to both spacetimes, then it is
sufficient if $\psi$ is a state on the algebra $\mathcal W
(\mathcal O)$. The restrictions of $B$ and $G$ to the region
$\mathcal O$ are Hadamard states on $\mathcal W(\mathcal O)$. Then
we can develop\footnote{The folia induced restrictions of $G$ and
$B$ to $\mathcal W(\mathcal O)$ coincide\cite{BFV}, which means
that we can always express, for instance, the expectation values
w.r.t. $G$ as expectation values with respect a state, say $\psi$,
described by a density matrix in the GNS Hilbert space induced by
$B$. Although none of these states is a ground state for $\mathcal
W(\mathcal O)$, it is not necessary (QEIs can be developed with
respect to arbitrary Hadamard states), as only their two-point
functions enter the derivation of  $Q_B[w,\vec x]$ \cite{Few_gen}
(see also the discussion of structure of QEIs in
\cite{FewRoman}).} QEI with respect to either of these states. For
instance, we have
\begin{equation}\label{as}
   \langle \p(w,\vec x)\rangle_B - \langle \p(w,\vec x)\rangle_G \geqslant Q_G[w,\vec
   x].
\end{equation}
Analogously, by changing the role of spacetimes used in the
argument, we will find
\begin{equation}
   \langle \p(w,\vec x)\rangle_G - \langle \p(w,\vec x)\rangle_B \geqslant Q_B[w,\vec x],
\end{equation}
from the quantum inequality derived with respect to $B$. As the
energy density of the Boulware state is known, the former of these
inequalities provides an upper bound on $\langle \p(f,\vec
x)\rangle_G$, while the latter provides a lower bound. In what
follows, we will investigate both of these bounds.

\subsection{Bound from below}
In this section we will recall the QEI w.r.t. the Boulware state
\cite{Few,Pfenning}, and show how it can be utilized to restrict
the energy densities of ground states for spacetimes of compact
objects. As we argue in appendix \ref{KMS}, the QEI \eqref{QEIss}
can be employed to restrict the difference of the energy densities
of the ground state $G$ and the Boulware state $B$ only if the
sampling function $\w^2(t)$ is such, that the smallest double cone
containing the part of world-line of the static observer, for
which \mbox{$w^2(t)>0$}, is contained in the region isometric with
the part of the Schwarzschild spacetime (see fig. \ref{shell}).

We will now review some of the properties of the operator $A$
specific to the Schwarzschild spacetime. The nature of the point
$r=\R$ (which is a regular singular point of the wave equation) is
similar to that of $r=\infty$ in this respect: none of the
generalized eigenfunctions of $A$ is locally square-integrable
near $r=\R$ (w.r.t. the scalar product \eqref{sc}). Kay has shown
\cite{kay}, that $A$ defined on $\dom(\Sigma_t)$ is essentially
self-adjoint, and therefore for a self-adjoint extension the
boundary conditions at $r=\R$ are imposed automatically. This
unique extension is such, that there remain two generalized
eigenfunctions of the operator $A$ for each positive eigenvalue
$\w^2$. They are usually denoted in the literature by $\Rup_{\w
l}(r)$ and $\Rin_{\w l}(r)$ (the former describes a wave purely
outgoing to $r=\infty$, while the latter a wave purely falling
through the horizon $r=\R$). Close to the horizon, an approximate
calculation due to Candelas \cite{Cand} reveals:
\begin{align}
\sum_{l=0}^\infty (2l+1)|\Rup_{\w l}(r)|^2&\approx\frac{4\w^2}{(1-\R/r)},\\
\sum_{l=0}^\infty (2l+1)|\Rin_{\w
l}(r)|^2&=\frac{1}{\R^2}\sum_{l=0}^\infty(2l+1)|\Bin_l(\w)|^2,\\
\dens{L}_B&\approx\frac{1}{480\, \pi^2 L^4}.\label{r_Boulware}
\end{align}
Although the sum involving $\Rin_{\w l}$ cannot be evaluated
analytically, Jensen, McLaughlin and Ottewill\cite{Ott1} have
found an explicit (though approximate) expression for the
transmission coefficients, $\Bin_l(\w)$, for
 $(\w\R\ll 1)$\footnote{More precisely, for $\R
\w\ll 1$ the generalized eigenfunctions of $A$ can be approximated
with the help of the method of asymptotic matching\cite{BO} (as
was discovered in \cite{Staro}, which yields a uniform
approximation of these functions (as $(\R\w)\ra 0$). A necessary
condition for asymptotic matching is the existence of an
intermediate region $D$, such that for all $x\in D$: $\R\w +1\ll
x\ll\frac{l+1}{\R\w}$. }.
\begin{equation}\label{otte}
  \Bin_l(\w)\approx \frac{(l!)^3}{(2l+1)!(2l)!}(-2i \w\R)^{l+1}
\end{equation}
In the case of long measurements, only the leading behavior of the
mode sums, for small frequencies $\w$ will be relevant; we will,
according to \eqref{otte}, simplify the second sum by retaining
the $l=0$ term only,
\begin{equation*}
\sum_{l=0}^\infty (2l+1)|\Rin_{\w l}(r)|^2\approx 4 \w^2
+\frac{1}{\R^2}\, O\left[(\w\R)^4\right].
\end{equation*}

Let us turn to the QEI with respect to the Boulware state, for an
observer located at the proper distance $L$ from the horizon (the
radial coordinate of this observer will simply be denoted by $r$).
We consider the re-scaled weight functions $w_\la(t)$ (see section
\ref{scaling}). The $Q_B[w_\la]$ \eqref{QEIss} for the QEI  with
respect to the Boulware state, after a further substitution $p\ra
\la p$, $\w\ra \la \w$, reads
\begin{multline}
  |Q_B[w,\vec x]|\approx \frac{1}{16\pi^3\la}\int_0^\infty \la d\w\int_0^\infty
  \frac{dp}{p}\\
  \left\{\frac{\la^2p^2}{1-\R/r}+\frac{1}{4r^2}\d_r \left[r^2 (1-\R/r)
  \d_r\right]\right\}\left[\frac{4\la^2p^2}{1-\R/r}+4\la^2p^2\right]\*
  g(\w+p).
\end{multline}
As the above expression is intended to be the leading term of the
asymptotic expansion of the exact $Q_B[w,\vec x]$ for $\la\ra 0$
we drop all the sub-dominant terms in $\la$, and  arrive at
\begin{equation*}
|Q_B[w,\vec x]|\approx \la^2\ \frac{\R^2}{16\pi^3 r^4
(1-\R/r)^2}\int_0^\infty d\w\ \int_0^\infty dp\ {p}\
g(\w+p)\approx (\la\R)^2 \frac{I}{\pi^3 L^4}\approx(\la
\R)^2\frac{480\, I}{\pi}|\dens{L}_B|,
\end{equation*}
where  an approximate relation between $r$ and $L$ (similar to
\eqref{a_rela}) was utilized, and the remaining, positive and
finite double integral of $g(p+\w)$ was denoted by $I$:
\begin{equation}\label{def_I}
I=\int_0^\infty d\w\ \int_0^\infty dp\ {p}\
g(\w+p)=\frac{1}{2}\int_0^\infty du\, g(u) u^2.
\end{equation}
Altogether, the following theorem was proved:

\begin{theorem}
   Let $G$ denote the restriction of the ground state on the
compact object's spacetime to the region isometric with the
Schwarzschild spacetime. The (time-independent) expectation value
 of the local energy density operator, evaluated at a
small distance $L$ from the horizon $\langle\p(L)\rangle_G$ is
bounded from below by:
\begin{equation}\label{bound_below}
    \langle\p(L)\rangle_G-\dens{L}_B\geqslant
    - (\la \R)^2\frac{480\,
I}{\pi}|\dens{L}_B|,
\end{equation}
where $\la$ is scale factor and $I$ is connected to $g(\w)$ and
$w(t)$ according to \eqref{def_I}, \eqref{def_g}. Note, that
$\dens{L}_B$ is given by \eqref{r_Boulware}.
\end{theorem}

{\it Remark:} If the sampling function $w^2(t)$ has a support on
an interval of unit length, then $\la=1/T$, where $T$  is the
rescaled time of measurement. The largest available $T$,
\eqref{T}, and therefore the smallest available $\la$, are the
only quantities in \eqref{bound_below} which depend on $\ell$,
that is, on the distance from the horizon up to which both
spacetimes are isometric. For given $\ell$, one can further
sharpen the bound by varying over allowed sampling functions $\w$
(i.e. by finding the minimal value of $I$, see appendix
\ref{opti}).

 We also note, that the bound \eqref{bound_below} can in principle
 be established at any distance $L$ from the horizon. However,
 only for small/large $L$ an explicit estimate of the
 mode sum is known \cite{Cand}, and therefore only in these cases
is it possible to approximate  the RHS of \eqref{bound_below} by
an explicit expression. Let us emphasize the physical meaning of
\eqref{bound_below}: in the limit  $\ell\ra 0$ one obtains from it
a strong result $\langle\p(L)\rangle_G\geqslant \dens{L}_B$.

\subsection{Remarks on the bound from above}\label{above}
In order to find a bound from above on $\dens{r}_G$, in the region
isometric with the Schwarzschild spacetime, we proceed as follows:
firstly we search for a QEI w.r.t. the state $G$; this inequality
allows all other states to exhibit only a limited negative energy
densities (negative means: below $\dens{r}_G$). Subsequently we
use the fact, that the energy density of the Boulware state is
known, and according to the QEI this density cannot be much lower
than $\dens{r}_G$. Such a reasoning provides an upper bound on
$\dens{r}_G$.

The mode sum of the generalized eigenfunctions of the operator
$A$, for the spacetime of a compact object, which is needed in
order to estimate $Q_G[w]$ \eqref{QEIss}, will not be known , even
approximately as in the case of Boulware state. Consequently, the
result we shall prove here will be weaker: we will only prove,
that there exists a QEI with respect to the ground state, and we
will estimate how the functional $Q_G[w,\vec x]$ decays for large
sampling times $T$.

To proceed, we note that as a consequence of the wave equation the
functional $Q_G[w,\vec x]$ can also be written in the form:
\begin{equation}\label{abs}
 Q_G[w,\vec x]\doteq-\frac{1}{2 \pi}\int_0^\infty
  d\w
  \int \frac{dp}{2p}\sum_{l,m} \left[\frac{p^2}{f(t)}\, |F_{I}(\vec x)|^2 -\d_i
  F_I(\vec x) \d^i \overline{F_I}(\vec x)
  \right]g(\w+p),
\end{equation}
from which it is evident that the integrand is pointwise positive.
We can now rescale the QEI, by taking the rescaled sampling
functions $w_\la(t)$. We prove in the appendix \ref{KMS} that
$Q_G[w_\la,\vec x]$ fulfills for small $\la$:
\begin{equation}
Q_G[w_\la,\vec x]<\la c_2\, Q_{KMS}[w_\la,\vec x],
\end{equation}
where $Q_{KMS}[w_\la,\vec x]$ is the form of the right-hand-side
of the QEI with a KMS state taken as the reference state. This
functional is bounded for each finite $\la$, and approaches a
finite limit for $\la\ra 0$ (at least for spacetimes with compact
Cauchy surfaces), as has been shown by Fewster and Verch
\cite{FV3}.

Although the above estimate constitutes
 an upper bound on the energy density of the ground
state $G$, its properties are not quite the same as that of the
lower bound (eq. \eqref{bound_below}). Not only is the assured
decay property (in general) weaker, but also we have no explicit
estimate (at present) of the absolute magnitude of the function
$c_2\* Q_{KMS}[w_\la,\vec x]$. Moreover, this estimate is not
uniform with respect to the changes of the internal geometry of
the object (i.e. it does still depend on $\ell$). Consequently,
even though we are able to prove the existence of an upper bound
on the energy density of the ground state (which is a new result),
its physical significance is weaker than that of
\eqref{bound_below} - in particular, it does not even assure, that
the energy density is negative, for sufficiently small $\ell$ at
fixed $L$. In order to derive a sharper bound, it appears
necessary to consider particular cases of concrete spacetimes, and
estimate the low-frequency behavior of the mode sums on these
spacetimes. We are currently pursuing this line of thought.

\section{Discussion and outlook}
In this work we have investigated the expectation values of the
energy density operator for ground states of a massless, scalar
quantum field, propagating on a static, spherically symmetric
spacetimes the exterior part of which is isometric to the
Schwarzchild spacetime.

We would like to recall the most important, in our opinion,
results of this paper: firstly, we have shown that the quantum
energy inequalities supplied with arguments taking into account
the necessary localization properties of observables, allow for a
development of meaningful bounds on the expectation value of the
energy density of a ground state, which is otherwise difficult to
compute analytically. This method was originally proposed by
Fewster and Pfenning in their investigation of the energy
densities for ground states on flat spacetime with boundaries
(Casimir setup) (\cite{FP}). Secondly, by means of a simple
geometrical observation, we have shown, that these bounds are
especially interesting if one considers curved-spacetime context
and the exterior region of the compact object stretches up to a
very small distance $\ell$ from the Schwarzschild horizon. In this
case, long times of measurement are granted (eq. \eqref{T}), and
these make the bounds provided by QEIs especially tight. Thirdly,
a concrete example of these bounds was developed, and it was
shown, that the energy density of a ground state can be more
negative than the energy density of the Boulware state (at the
same point) only by a limited amount, with the bound depending
significantly on $\ell$ (the difference falls as $1/\ln^2(\ell)$).
Finally, we have made some preparatory steps for obtaining an
upper bound on the energy density of the ground states. In
particular: by recalling some abstract results on the existence of
ground and KMS (thermal equilibrium) states, for a class of
spacetimes fulfilling the condition \eqref{condition}, we have
argued, that there will exist a bound of similar nature (provided
by the QEI), and we have estimated the necessary rate at which
this bound ``sharpens'' for long times of measurement. Moreover,
we have shown, that only the low-frequency (that is $\w\R\ll 1$)
behavior of the ``mode-sums'' will be necessary for a development
of a concrete upper bound on the energy density of the ground
state. We hope to return to the issue of upper bound in a future
publication.

  It should be clear, that if the bound from above, the existence of which we have proved here,
turned out in a more careful analysis to be of similar nature to
the bound from below (established here), then the physical
consequences would be profound. One could argue, by investigating
ground states for a family of static spacetimes with the value of
$\ell$ decreasing towards zero, that quantum field-theoretical
effects necessarily become significant in this case, that is the
energy density of ground states would necessarily be large in
magnitude and negative (cf. \cite{Laughlin} for a physically
motivated proposal in this direction). It is precisely this type
of behavior that Roman and Bergmann found \cite{RB} as
indispensable in their general search for energy densities
necessary for reverting the formation of trapped surfaces. Note,
however, that what they found as physically implausible, namely,
that the energy conditions need to be violated in the region of
low density of ordinary matter, would become a necessity - as a
consequence of the bound from above, should a bound with
properties similar to the bound from below indeed be found.

With respect to the generalization of the results of this paper we
note, that the energy density also diverges to minus infinity for
a number of ground states of static spacetimes with horizons
\cite{BD}, notably for the static parametrization of a part of the
de Sitter spacetime (which is of interest for scenarios presented
in \cite{MM,Laughlin,VW}). It appears, that our arguments can be
generalized to these spacetimes as well (with the Schwarzschild
spacetime replaced by another static spacetime with a horizon). We
remark, that our bounds can also be employed as a ``consistency
check'' for approximate or numerical computations of energy
densities for spacetimes of compact objects (such as these
presented in \cite{Anderson}): typically in such cases one has an
approximate version of the two-point function as well as the
energy density. The low-frequency behavior of the former can be
employed in order to derive a bound, which must be satisfied by
the latter.

The most far-reaching goal which we see for the type of arguments
presented here would be to remove the assumption that the
spacetime is static and the field resides in the ground state, and
instead investigate the energy densities for spacetimes undergoing
a (spherical) gravitational collapse. It would be particularly
interesting to take a spacetime, which was initially static (and
therefore possessed a distinguished ground state which could
followed for later times), and later collapsed to a Schwarzschild
black hole. For such a situation Fredenhagen and Haag determined
in a rigorous manner \cite{FH}, that the inward-looking detector
(at late times and large distances from the horizon) registers the
Hawking radiation. If their analysis could be adapted to detectors
located close to the horizon, or equivalently - if one could argue
that the state investigated by these authors  approximates the
Unruh vacuum state at late times, then this would give all the
information needed for a quantum inequality w.r.t. the
non-stationary state of quantum fields on the spacetime of a
collapsing object (for the Unruh vacuum there are approximate
expressions for the two-point function, as well as the
energy-density \cite{Cand}).

\begin{acknowledgments}
I would like to thank Chris Fewster for his critical remarks,
particularly in the early stages of this work, as well as for
making the results of his collaborative work with Michael Pfenning
available to me prior to publication.
\end{acknowledgments}

\appendix
\section{Localization of non-linear observables for QFT in curved spacetimes}
\label{local} For our arguments in this paper it is essential,
that there exists the largest allowed amount of time for a local
common measurement made by (static) observers located at the
position corresponding to the distance $L$ from the horizon.
(Otherwise, the bounds provided by quantum inequalities - for
infinite time of measurement - would imply equality of the energy
densities $\dens{L}_B$ and $\dens{L}_G$). This restriction on the
time of measurement is provided by the the necessity to fit the
smallest double-cone containing the observer's world-line during
the measurement to the region, where the spacetimes of interest
are isometric (the exterior region). On general grounds, taking
into account the hyperbolic character of the underlying wave
equation, it is known that local observables (even linear fields)
belong to the algebras associated with causally complete regions
(see chapter III.3 of \cite{Haag}, and the time-slice axiom in
\cite{BFV}), and double cones are examples of such regions. Here
we would like to point out that this localization is not only a
formal requirement, but rather a physical necessity. This
corroboration stems its relation to the
 principle of ``Local Position Invariance'', which is one of
the forms of the equivalence principle\footnote{The discussion of
the equivalence principle (which has various ingredients) as well
as its current experimental status can be found in an excellent
review article by C.Will \cite{Will}.}. It asserts that  \emph{the
outcome of any non-gravitational experiment is independent of
where and when in the universe it is performed; the fundamental
constants of non-gravitational physics should be constants in
space and time.}

\begin{figure}[h]\centering
\includegraphics[scale=0.8]{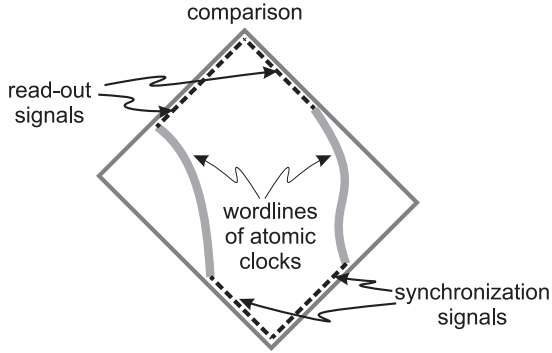}
\caption{GPS-like test of local position invariance. The
experimental result is allowed to depend only on the gravitational
field in the smallest causal normal neighborhood containing the
whole measurement apparatus (together with the final signal
read-out).} \label{gps}
\end{figure}

In experiments the LPI is verified with the help of the following
procedure (see figure \ref{gps}): two precise frequency standards
(such as atomic clocks) are synchronized with a light signal. They
follow their world-lines (which can assumed to be geodesic lines
in order to rule out the influence of acceleration on the clocks)
and continuously send light signals which carry the information
about their states. Those signals are compared at a single event.
The result is scrutinized against general relativistic
predictions, namely, one calculates the geometric lengths of both
world-lines. If the time lapse the atomic clocks have measured is
proportional to the length of their world-lines, then indeed the
non-gravitational experiments (here the quantum optical
experiments) are independent of the position in the universe.
Therefore, clearly, if the LPI is fulfilled then the experimental
result depends only on the geometry in the region of spacetime
which contains world-lines of the clocks and all the causal
geodesics which join them (the clocks must be synchronized and
their state must be compared). Thus, the LPI which at the moment
is supported by strong experimental evidence \cite{Will} implies
that the results of experiments depend on the
 geometry in the double-cone containing the entire
measurement setup. This is precisely the localization region for
observables, which we use in this paper.

\section{Decay of QEIs for large sampling times and the existence of ground and
thermal equilibrium states}\label{KMS}

 In this section we will argue, that the spacetimes of (certain)
compact, spherical objects admit ground and KMS (thermal
equilibrium) states for massless fields. The existence of a ground
state is an a priori condition for an investigation of the
expectation value of the energy density operator with respect to
this state, which we do throughout this paper. On the other hand,
the existence of KMS states provides a general estimate of the
long term  behavior of the QEI, which is necessary in section
\ref{above}.

In the part concerned with the existence of ground and KMS states
will use the approach and the methods developed by Kay \cite{kay},
and Kay and Wald \cite{KW}, necessary for the construction of
ground and KMS states on the Schwarzschild spacetime (exterior
part of the Kruskal spacetime).

In the case of Schwarzschild spacetime Kay noted \cite{kay}, that
although $A$ does not have a positive lower bound (for massless fields),
it fulfills
\begin{equation}\label{kay_bound}
  \int_{\Sigma_t} u\ Au\ dx d\Vac\geqslant \int_{\Sigma_t}\a_S(x) |u|^2\ dx d\Vac,
\end{equation}
for $u\in\dom(\Sigma_t)$, with a strictly positive function $\a_S(x)$:
\begin{equation}\label{cond_Kay}
\a_S(x)=\frac{g\d_r g}{r}=\frac{\R(1-\R/r)}{r^3},
\end{equation}
(the index $S$ indicates, that $\a_S$ is related to the Schwarzschild
spacetime).  This property allowed him to conclude that

\begin{lemma}[Proposition A4.9 and Theorem 4.5 of
\cite{kay}]\label{lemat} In the case of Schwarzschild spacetime,
the operator $A$ is such, that
\begin{itemize}
  \item $\dom(\Sigma_t)\subset \mathcal D ({\a_S}^{-1/2})\subset \mathcal D
  ({A}^{-1/2})$, where $\mathcal D(A^{-1/2})$ denotes the domain of the
  inverse of the square root of $A$, which coincides with $\mathcal D(\h^{-1})$.
  \item The one-particle structure $(k,\mathcal H, e^{i\h t})$ is
  regular over the symplectic space $(S,\sigma,\mathcal V(t))$, $S=\dom(\Sigma_t)\oplus
  \dom(\Sigma_t)$. This means that $\h$ is strictly positive, and
  $k  S\subset \mathcal D(\h^{-1/2})$.
\end{itemize}
\end{lemma}

The results of Kay, quoted above, can be generalized to the case of
static spherically symmetric spacetimes of (some) compact objects. For
these spacetimes, let us introduce the condition
\begin{equation}\label{condition}
\a(r)=\frac{g\d_r g}{r} > 0 \qquad{\forall r\in[0,\infty)},
\end{equation}
which due to $g>0$ reduces to $\d_r g(r)>0$. Using standard
notation for spherically symmetric spacetimes generated by a
perfect fluid \cite{Wald},
\begin{equation}
    m(r)=4\pi \int_0^r ds\, \rho(s) s^2,
\end{equation}
we find that condition \eqref{condition} is equivalent to
\begin{equation}
    m(r)+2\pi r^3[p(r)-\rho(r)]>0.
\end{equation}
For the above condition to be fulfilled it is sufficient (although
not necessary), if the spacetime is generated by a fluid of
positive pressure with positive density which is a non-increasing
function of $r$, \footnote{Obviously,  for such spacetimes $\ell$
cannot be small compared to $\R$ \cite{Wald}.}.

For spacetimes fulfilling the condition \eqref{condition} an
analogue of the theorem of Kay can be proven:

\begin{lemma}For spherically symmetric, static spacetimes, whose metric is parameterized as in
\eqref{metric}, which satisfy \eqref{condition}, the conclusions
of the lemma \ref{lemat} hold.
\end{lemma}
\begin{proof} The will adapt the proof of Kay \cite{kay}, which was established
for the Schwarzschild spacetime. The essential spacetime-specific
feature which he uses, is the condition \eqref{kay_bound} together
with \eqref{cond_Kay}. They are replaced here, by assumption, by
the condition \eqref{condition}. Apart from this, we do not see
any part of the proof which would need to be altered (note, that
our Cauchy surfaces $\Sigma_t$ are complete by assumption and
therefore the standard arguments employing the well-posedness of
the Cauchy problem, which imply essential self-adjointness of $A$
on $\dom(\Sigma_t)$, can also be applied).
\end{proof}

With these statements at hand we may also conclude (again for
spacetimes fulfilling \eqref{condition}) that the KMS (thermal
equilibrium) states exist, due to the theorem of Kay and Wald:
\begin{lemma}[Lemma 6.2 of \cite{KW}] If the one-particle ground state
structure is regular (i.e. if the Hamiltonian $\h$ is strictly
positive), and if\
\begin{equation}k S\subset \mathcal
D(\h^{-1/2}),
\end{equation}
 then the KMS (thermal) states can be constructed for
any inverse temperature $\beta\in \mathbb R^+$. The symplectic
space $S$ and the operator extracting positive frequencies, $k$,
are defined as in section \ref{pre}.
\end{lemma}

The asymptotic behavior of QEIs derived with respect to the ground
state can be approximated if the KMS states on the considered
spacetime exist. Let us recall the form of the KMS two-point
function, smeared in both time variables (separately) with a real,
compactly supported function $\chi$:
\begin{equation}
\lim_{\vec y\ra \vec x}\w^\be_2(\chi,\vec x,\chi,\vec
y)=\int_0^\infty \frac{d\w}{2 \w}\sum_{l,m}\ \coth(\be \w/2) |\hat
\chi (\w)|^2 |F_{lm}(\w,\vec x)|^2,
\end{equation}
By existence of KMS states we know, that the integrand is locally
integrable near $\w=0$. In chapter IV, it was necessary to
estimate the long-term (that is: small $\la$) behavior of the
right-hand-side, $Q_G[w_\la,\vec x]$, of quantum inequalities
derived with respect to the ground state $G$, \eqref{abs}. Clearly
it is the term with derivatives that induces the dominant
behavior; we will estimate it here. For sampling functions $w(t)$
such, that $\hat w(p)$ is sufficiently concentrated around $p=0$
there exist constants $c_1$ and $c_2$ such that the dominant part
of $Q_G[w_\la,\vec x]$ fulfills (with $\p=\la p'$, $\w=\la w'$)
\begin{align*}
\la &\int_0^\infty  d\w\int_0^\infty \frac{dp}{p}\sum_{lm}\,
\frac{1}{\la} |\d_i F_{lm}(\w, \vec x)|^2 g_\la(\w+p)=\\ =\la
&\int_0^\infty  d\w'\int_0^\infty \frac{dp'}{p'}\sum_{lm}\,
\frac{1}{\la^2} |\d_i F_{lm}(\w'\la, \vec x)|^2 g(\w'+p')<\\
<\la\, c_1 & \int_0^\infty  d\w'\int_0^\infty
\frac{dp'}{p'}\sum_{lm}\, \frac{1}{\la^2\, p'} |\d_i
F_{lm}(\w'\la, \vec x)|^2 g(\w'+p')=\\ =\la\, c_1 &\int_0^\infty
d\w\int_0^\infty \frac{dp}{p^2}\sum_{lm}\,  |\d_i F_{lm}(\w, \vec
x)|^2 g_\la(\w+p)<\\< \la\, c_2 &\int_0^\infty d\w\int_0^\infty
\frac{dp}{p}\sum_{lm}\, \coth(\be \w/2) |\d_i F_{lm}(\w, \vec
x)|^2 g_\la(\w+p).
\end{align*}
But the latter is just the form of $Q_{KMS}[w_\la,\vec x]$,
multiplied with $\la c_2$, that would be obtained for a quantum
energy inequality with the KMS state as the reference state: in
particular it is bounded for each finite $\la$. It is remarkable,
that the limit $\la\ra 0$ of $Q_{KMS}[w_\la,\vec x]$ exists (is
finite). This has been shown by Fewster and Verch in their paper
on the relation between QEIs and thermodynamic equilibrium
conditions, \cite{FV3}. This is the conclusion of their theorem
4.7, which asserts that passive (in particular KMS) states fulfill
what they call a \emph{limiting} QEI. It follows, that the
following limit
\begin{equation}
\lim_{\la\ra 0}
    \int_0^\infty d\w\int_0^\infty \frac{dp}{p}\sum_{lm}\, {\coth(\be
\w/2)} |\d_i F_{lm}(\w, \vec x)|^2 g_\la(\w+p)
\end{equation}
exists. Although we will always have a finite $\la$ (due to the
locality requirement), this result shows that our estimate picks
up the leading term of the asymptotic expansion of $Q_G[w_\la,\vec
x]$.  However, it should be mentioned, that Fewster and Verch have
assumed compactness of the Cauchy surfaces $\Sigma_t$, and that
this assumption appears essential to their derivation. Our
estimate of the decay of $Q_G[w_\la,\vec x]$ can be seen as a
strengthening of their result for ground states, namely, it
asserts that if KMS states exist on a static spacetimes with
compact Cauchy surfaces, then the right-hand-side of the QEI with
respect to the ground state decays at least as $\la^{-1}$ for
large sampling times (small $\la$).

\section{Note on the optimization of quantum energy
inequalities} \label{opti} The lower bound on the energy density
of ground states developed in this paper, \eqref{bound_below},
still contained a possibility to optimize among sampling functions
$w^2(t)$ of compact support in the interval of unit length. More
precisely, the optimization problem can be formulated as follows:
one asks for a minimal value of the functional
\begin{equation}
    I=\frac{1}{2}\int_0^\infty du\, |\hat w(u)|^2 u^2,
\end{equation}
for real functions $w(t)$, which are compactly supported in the
interval $[0,1]$, with the normalization condition $\int_0^1 dt\,
|w(t)|^2=1$. The functions in the domain of $I$ need not to be
smooth (i.e. it suffices if they lead to finite values of the
functional). If we note that for symmetric functions (i.e.
$w(t)=w(1-t)$)
\begin{equation}
    I={\pi} \int_{-\infty}^\infty \left|\frac{dw}{dt}\right|^2,
\end{equation}
then the optimization problem can easily be solved, namely the
functions $w(t)$ must necessarily vanish at the endpoints of the
interval\footnote{For functions which are not symmetric, the same
argument applies:  they do not belong to the domain of the
functional $I$ because of the singularity of $dw/dt$ at the
endpoints.}, and therefore can be expanded in the series of
eigenfunctions of the (selfadjoint) operator
\begin{equation}
    H=-\frac{1}{2}\frac{d^2}{dt^2}
\end{equation}
with the Dirichlet boundary conditions at the endpoints. Thus the
optimization problem reduces to a simple quantum mechanics: the
normalization condition for $w(t)$ is easily seen as the usual
normalization of wave functions, and the functional $I$ is just
the expectation value of $H$ in a state described by $w(t)$.
Obviously the eigenstate corresponding to the lowest eigenvalue,
$w_0(t)=\sqrt{2} \sin(\pi t)$, minimizes such a functional.
Therefore the minimal value of $I$ is $\pi^2/2$. We note, that a
similar argument can be used for less-trivial optimization
problems turning up in the context of QEIs, namely one can apply
them for functionals of the form \mbox{$I_\a=\int_0^\infty du\,
|\hat w(u)|^2 u^\a$} with $1\leqslant \a \leqslant 2$, as in this
case the domain of the Dirichlet extension of $H$ is dense in the
domain of $I_\a$.

\end{document}